\RequirePackage{rotating}
\documentclass[useAMS,usenatbib]{mn2e}
\usepackage{epsfig,graphicx,aas_macros,subfig,float}
\usepackage{graphics,rotating}
\usepackage{amsmath}

\begin{document}
\title[Orbital evolution and eccentricity in 4U 1700--37]{Orbital evolution and search for eccentricity and apsidal motion in the 
eclipsing HMXB 4U 1700--37}
\date{}
\author[Nazma Islam and Biswajit Paul]{Nazma Islam$^{1,2}$\thanks{E-mail:
nazma@rri.res.in;} and Biswajit 
Paul$^{1}$ \\
$^{1}$Raman Research Institute, Sadashivnagar, Bangalore-560080, India\\
$^{2}$Joint Astronomy Programme, Indian Institute of Science, Bangalore-560012, India}

\maketitle

\begin{abstract}

In the absence of detectable pulsations in the eclipsing High Mass X-ray binary 4U 1700--37, the orbital period decay is necessarily determined from the 
eclipse timing measurements. We have used the earlier reported mid-eclipse time measurements of 4U 1700--37 together with the new measurements from long term 
light curves obtained with the all sky monitors {\it RXTE}--ASM, {\it Swift}--BAT and {\it MAXI}--GSC, as well as observations with {\it RXTE}--PCA, to 
measure the long term orbital evolution of the binary. The orbital period decay rate of the system is estimated to be 
${\dot{P}}/P = -(4.7 \pm 1.9) \times 10^{-7}$ yr$^{-1}$, smaller compared to its previous estimates. We have also used the mid-eclipse times and the 
eclipse duration measurements obtained from 10 years long X-ray light-curve with {\it Swift}--BAT to separately put constraints on the 
eccentricity of the binary system and attempted to measure any apsidal motion. 
For an apsidal motion rate greater than 5 degrees per year, the eccentricity 
is found to be less than 0.008, which limits our ability to determine the apsidal motion rate from the current 
data. We discuss the discrepancy of the current limit of eccentricity with the earlier reported values from radial velocity measurements of the companion star.

\end{abstract}

\begin{keywords}
X-rays: binaries - X-rays: individual: 4U 1700--37 - stars: binaries: eclipsing

\end{keywords}

\section{Introduction}

\subsection{Orbital evolution and apsidal motion in X-ray binaries}

The orbits of X-ray binaries evolve due to various mechanisms like mass and angular momentum exchange between the compact object and the companion star, 
tidal interaction between the binary components \citep{lecar1976,zahn1977}, magnetic braking \citep{rappaport1983}, stellar wind driven angular momentum 
loss \citep{brookshaw1993,heuvel1994}, X-ray irradiated wind outflow \citep{ruderman1989} and gravitational wave radiation \citep{verbunt1993}. 
In addition to orbital period evolution, the elliptic orbits of X-ray binaries also undergo apsidal motion. The classical apsidal motion is caused by 
tidal force \citep{cowling1938,sterne1939} and hence the rate of apsidal angle change is directly related to the stellar structure constant of the 
component stars \citep{kopal1978,claret1993}. For an accreting X-ray pulsar, repeated measurements of orbital parameters by pulse timing analysis at 
separate intervals of time is an efficient way to study the orbital evolution of the binary system  (Cen X--3 -- \citealt{kelley1983}; 
Her X--1 -- \citealt{staubert2009}; SMC X--1 -- \citealt{levine1993}; LMC X--4 -- \citealt{levine2000,naik2004}; 4U 1538--52 -- \citealt{mukherjee2006}; 
SAX J1808.4--3658 -- \citealt{jain2008}; OAO 1657--415 -- \citealt{jenke2012}) as well as its rate of apsidal motion (4U 0115+63 -- \citealt{raichur2010a}). 
For eclipsing HMXB pulsars like Vela X--1 \citep{deeter1987} and 4U 1538--52 \citep{falanga2015}, the rate of apsidal motion can also be calculated from the 
offset in the local eclipse period and the mean sidereal period, which is determined from pulse timing analysis. In case of eclipsing X-ray binaries, 
eclipse timing technique is used to determine the orbital evolution of binary systems (EXO 0748--676 -- \citealt{parmar1991,wolff2009};  
4U 1822--37 -- \citealt{jain2010}; XTE J1710--281 -- \citealt{jain2011}) as well as estimating parameters of the companion star and masses of the compact 
object \citep{coley2015,falanga2015}. In eclipse timing technique, the mid-eclipse times are used as fiducial markers to study any change in 
orbital period of the binary system. Mid-eclipse timing measurements have also been used to determine the rate of apsidal motion and other orbital 
parameters in case of eccentric optical eclipsing binaries \citep{gimenez1983,wolf2004,zasche2014}. In the absence of pulsations or eclipses, stable 
orbital modulation curves have also been found useful for measurements of orbital evolution of some X-ray binaries like Cyg X--3 \citep{singh2002} and 
4U 1820--30 \citep{peuten2014}. 
\par
Orbital decay of the compact HMXBs are also of interest in the context of short GRBs and gravitational wave astronomy as these are the progenitors of 
double compact binaries. The massive companion will leave behind a second compact star and if the two compact stars survive as a double compact binary, 
eventual merging of the two stellar components are believed to produce the short GRBs and are also expected to produce the sources for gravitational 
wave detection \citep{belczynski2002,abadie2010}.

\subsection{4U 1700--37}
The massive X-ray binary 4U 1700--37 was discovered with {\it Uhuru} \citep{jones1973}, which revealed it to be an eclipsing binary 
system with an orbital period of 3.412 days. The optical companion, HD 153919, is a O6.5 Iaf+ star, situated at a distance of 1.9 kpc and one of the 
most massive and hottest stars known in an HMXB \citep{ankay2001}. The nature of the compact object is uncertain due to lack of X-ray pulsations from the 
system \citep{rubin1996}. Lack of pulsations from 4U 1700--37 makes it difficult to determine the parameters of the binary orbit, especially eccentricity of the orbit and 
hence the rate of apsidal motion. Estimation of the orbital parameters using radial velocity measurements of the companion star HD 153919 from the 
ultraviolet and optical spectral lines \citep{hutchings1974,heap1992,clark2002,hammerschlag2003} was complex due to extreme mass loss rate of the 
companion as well as very high stellar wind. Previous measurements of orbital parameters by \cite{hammerschlag2003} using ultraviolet spectral lines 
with {\it International Ultraviolet Explorer}, found it to satisfy an orbital solution with an eccentricity {\it e} $\sim$ 0.22. 4U 1700--37 is an 
archetypal system to study the orbital evolution using eclipse timing as it is a bright source with sharp eclipse transitions in the hard X-rays.
\par
Earlier measurements of the mid-eclipse times spanning 20 years, both from the single pointed observations covering one full orbital cycle from 
{\it Uhuru} \citep{jones1973}, {\it Copernicus} \citep{branduardi1978} and {\it EXOSAT} \citep{haberl1989} as well as continuous observations with the 
all sky monitors {\it Granat}/WATCH \citep{sazonov1993} and {\it BATSE} \citep{rubin1996}, showed an orbital period decay rate of 
${\dot{P}}/P = -3 \times 10^{-6}$ yr$^{-1}$ \citep{rubin1996}. 
Recent work by \cite{falanga2015}, using the mid-eclipse measurements from {\it RXTE}--ASM and {\it INTEGRAL}, along with the previous mid-eclipse measurements 
from \cite{rubin1996}, showed a slower orbital decay with a rate of ${\dot{P}}/P = -1.6 \times 10^{-6}$ yr$^{-1}$. 
\par
In this work, we have used the above mentioned mid-eclipse times and the new mid-eclipse time measurements to obtain a long term eclipse history and 
orbital evolution of the system. We then investigated the possibility of estimating or constraining the eccentricity and the rate of apsidal motion of 
4U 1700--37 using the mid-eclipse timing measurements and the eclipse duration measurements independently with 10 years of {\it Swift}--BAT light curves.

\begin{figure*}
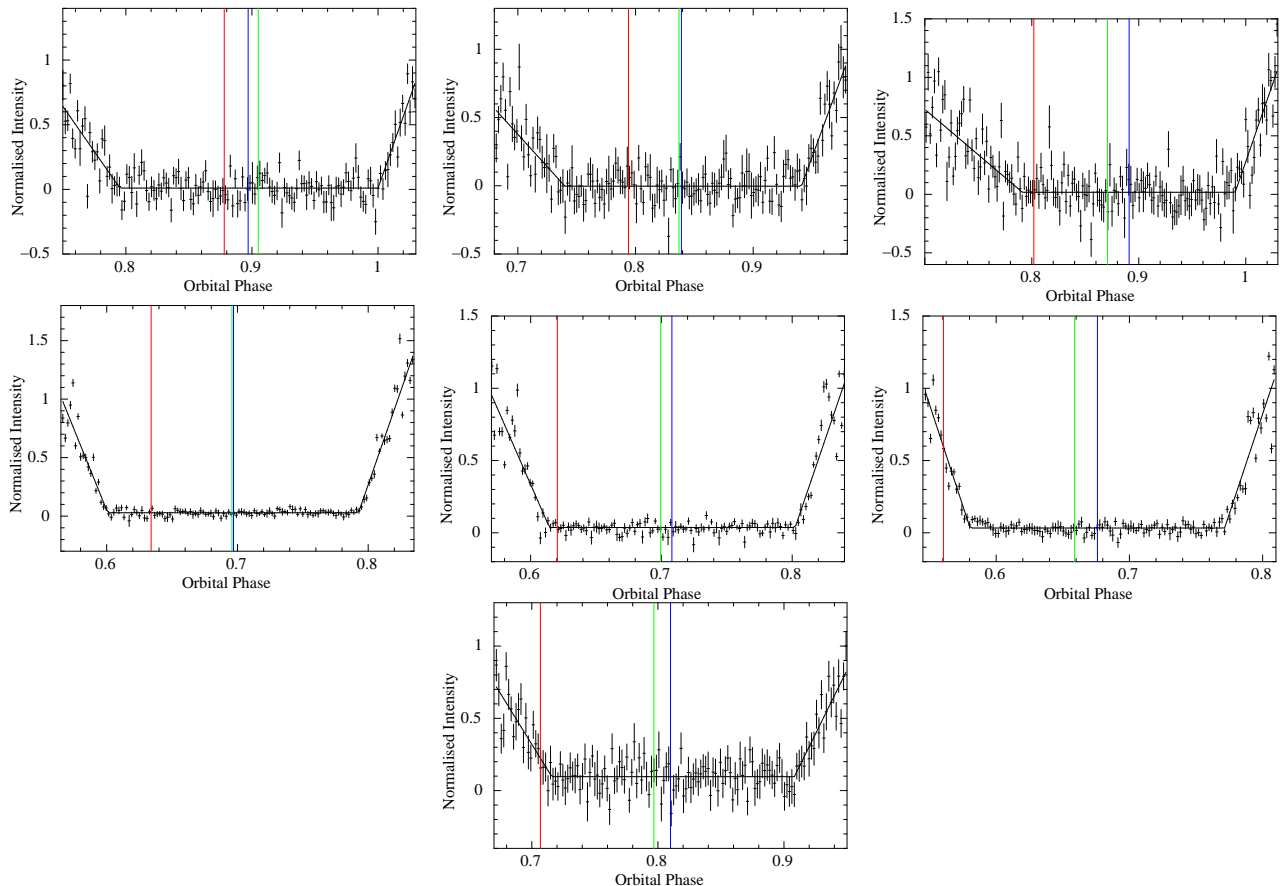

\centering
\includegraphics[angle=-90,scale=0.22]{seg1_asm_replot.ps} 
\includegraphics[angle=-90,scale=0.22]{seg2_asm_replot.ps} 
\includegraphics[angle=-90,scale=0.22]{seg3_asm_replot.ps} \\
\includegraphics[angle=-90,scale=0.22]{seg1_bat_replot.ps} 
\includegraphics[angle=-90,scale=0.22]{seg2_bat_replot.ps} 
\includegraphics[angle=-90,scale=0.22]{seg3_bat_replot.ps} \\
\includegraphics[angle=-90,scale=0.22]{maxi_replot.ps} 
\caption{Orbital intensity profile near the eclipse constructed for 3 light curve segments for 5-12 keV {\it RXTE}--ASM (top panel), 3 light curve segments for 15-50 keV 
{\it Swift}--BAT (middle panel) and 5-20 keV {\it MAXI}--GSC light curve (bottom panel). Blue line denotes the measured mid-eclipse phase, red line 
denotes the mid-eclipse phase expected from an orbital period change of ${\dot{P}}/P = -3.3 \times 10^{-6}$ yr$^{-1}$ from Rubin et al. (1996) and green line is the 
expected mid-eclipse phase from an orbital period change of ${\dot{P}}/P = -1.6 \times 10^{-6}$ yr$^{-1}$ from Falanga et al. (2015).}
\label{asm}
\end{figure*}

\section{Orbital Period Evolution}

\subsection{Mid-eclipse time measurements}
We have used the earlier reported mid-eclipse measurements from single pointed observations covering the X-ray eclipse with {\it Uhuru} \citep{jones1973}, 
{\it Copernicus} \citep{branduardi1978} and {\it EXOSAT} \citep{haberl1989}. We have also used the mid-eclipse time from a {\it Copernicus} observation 
in 1974 \citep{mason1976} which was not used in the previous studies because it showed higher residuals in the linear fit to the mid-eclipse 
times \citep{rubin1996,falanga2015}. However, as it will be shown later in Figure \ref{delay1}, this datapoint is less than 2 $\sigma$ different from the 
quadratic fit and is not an outlier anymore (blue datapoint in Figure \ref{delay1}).
Mid-eclipse times were also reported from 
the long term observations with X-ray all sky monitors {\it Granat}/WATCH \citep{sazonov1993}, {\it BATSE} \citep{rubin1996} and {\it INTEGRAL} 
\citep{falanga2015}, which are also used in the present analysis. 
\par
\begin{figure*}
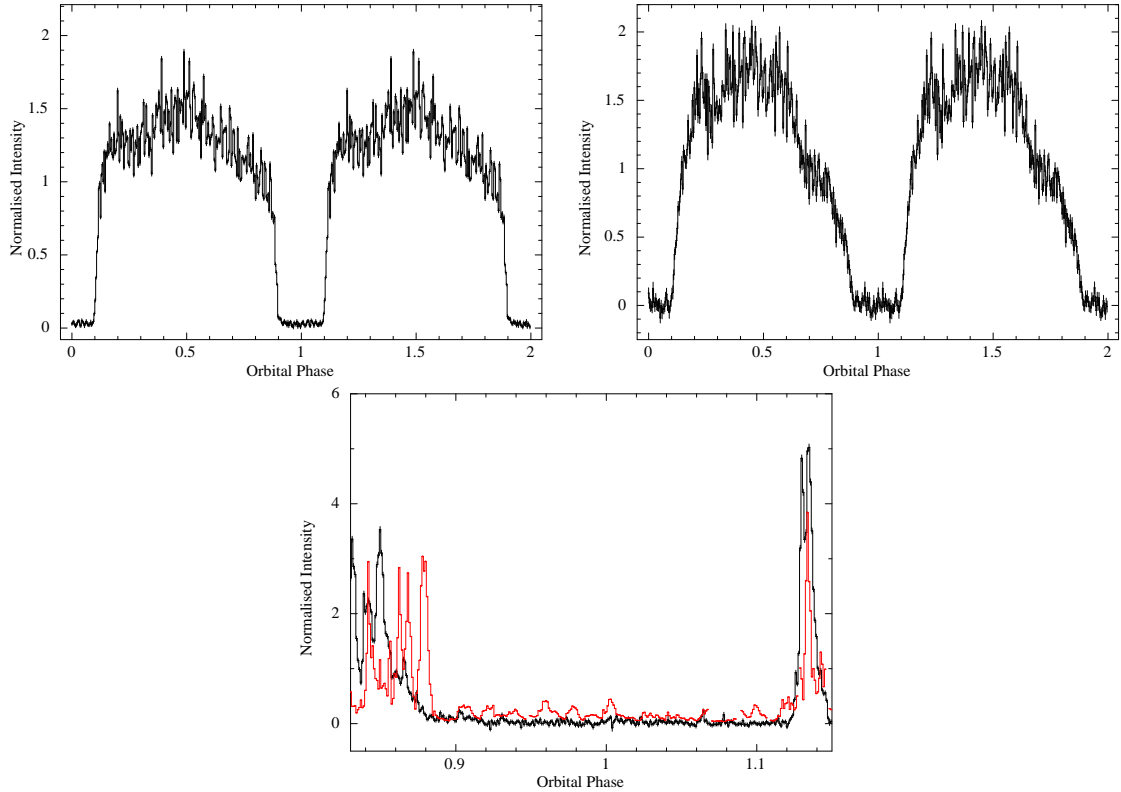

\centering
\includegraphics[angle=-90,scale=0.3]{swift_bat.ps} 
\includegraphics[angle=-90,scale=0.3]{rxte_asm_ch3.ps}
\includegraphics[angle=-90,scale=0.3]{overlaid.ps} 
\caption{{\it Top panel}: Plot of the orbital intensity profile created out of 10 years of {\it Swift}--BAT data in 15-50 keV energy band (left panel) and 16 years of 
{\it RXTE}--ASM in 5-12 keV energy band (right panel), which shows an averaged smooth profile with sharp eclipse. {\it Bottom panel}: Plot of the folded orbital intensity 
profile near the eclipse using {\it RXTE}--PCA (red line) observations overlaid on the {\it EXOSAT}--GSPC orbital profile (black line), which shows the presence of flares and 
dips just outside the eclipse.}
\label{gspc_pca}
\end{figure*}
\begin{figure*}
\centering
\includegraphics[angle=-90,scale=0.45]{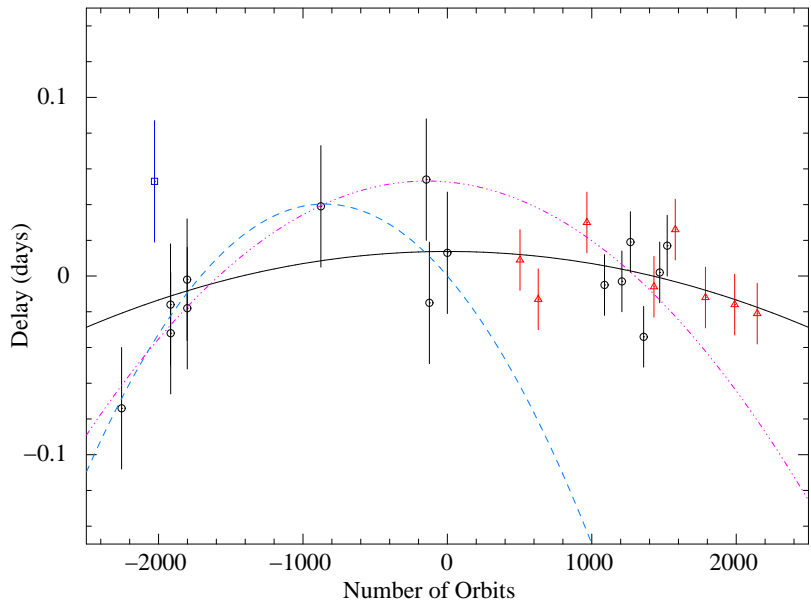} 
\caption{Delay in mid-eclipse times with respect to a constant orbital period. The black circles are the archival measurements of the mid-eclipse times and 
the red triangles are the new measurements of mid-eclipse times. The blue box datapoint is the mid-eclipse time measurement from {\it Copernicus} 1974 observation, which was 
not used in previous studies on the orbital evolution of the system but is used in this study. The errors on the mid-eclipse times are re-calculated as mentioned in 
Section 2.1. The solid black line is the quadratic component of the best fits to the mid-eclipse times. The blue dashed line and the magenta dot-dashed line are the 
quadratic components of the best-fits reported in Rubin et al. (1996) and Falanga et al. (2015) respectively.}
\label{delay1}
\end{figure*}

We have used long term light curves from {\it RXTE}--ASM \citep{levine1996}, {\it Swift}--BAT Transient Monitor \citep{krimm2013} and 
{\it MAXI}--GSC \citep{matsuoka2009} to determine new mid-eclipse times of 4U 1700--37. 
\cite{falanga2015} had also reported mid-eclipse times from {\it RXTE}--ASM lightcurves in the energy band of 1.5--12 keV. However, as seen in Figure 2 in 
\cite{falanga2015}, the eclipse is sharper and more pronounced in the light-curve of {\it RXTE}--ASM in 5--12 keV energy band compared to its 
profile seen in 1.3--3 keV and 3--5 keV energy band. This is due to photo-electric absorption of X-rays by the dense stellar wind of the companion which affects the soft 
X-rays photons more than the hard X-rays photons \citep{meer2007}. Therefore, we have used 16 years long 5--12 keV light curve with {\it RXTE}--ASM in this 
work and divided it into 3 segments, 
each consisting of about 5 years of data. The 10 years of {\it Swift}--BAT light curve in the 15-50 keV energy band was divided into 3 segments, each of duration 3.3 years. 
The {\it MAXI}--GSC light curve for 4.5 years was extracted in 5-20 keV energy band using the {\it MAXI} on demand data processing\footnote{http://maxi.riken.jp/mxondem/}. 
We have applied barycentric corrections to the light-curves using the {\small FTOOLS} task `earth2sun'. 
Eclipses are barely identifiable in only few segments of {\it Swift}--BAT light-curves, which has the highest sensitivity amongst the three X-ray 
all sky monitors in hard X-rays utilized here. In addition, the instruments have very sparse and uneven sampling making it impossible to measure the 
mid-eclipse times from unfolded light-curves which requires both ingress and egress to be sampled adequately, necessitating folding of light-curves.
The orbital period of the system is estimated separately for each segments 
of light curves using the {\small FTOOLS} task `efsearch' and then these light-curves are folded with their respective orbital period with the {\small FTOOLS} task `efold' 
to create orbital intensity profiles. The orbital intensity profiles near the eclipse were fitted with two ramp function having different linear ingress 
and egress profiles and a constant count-rate during the eclipse, given in Equation ~\ref{ramp}
\begin{eqnarray}
\label{ramp}
F(X)         &=& P3  \quad \mathrm{for} \quad (P1-P2) < X < (P1+P2) \nonumber \\
             &=& P4 \times (X-(P1+P2)) + P3 \quad \mathrm{for} \quad X > (P1+P2) \nonumber \\
             &=& -P5 \times (X-(P1-P2)) + P3 \nonumber \\
             & & \mathrm{for} \quad X < (P1-P2) 
\end{eqnarray}
where P1 is the mid-eclipse phase, P2 is the half width of the eclipse, P3 is the count-rate during eclipse, P4 is the slope during eclipse egress, 
P5 is the slope during eclipse ingress. The fitting of the function is done in the orbital phase range of about $\pm$ 0.15 around the mid-eclipse phase. 
The orbital intensity profile outside the eclipses have not been used for fitting with the double ramp function as it is energy dependent 
and has presence of flares outside eclipse as seen in the top panels of Figure \ref{gspc_pca}. The inverse of the slope of the eclipse ingress and 
eclipse egress are an approximate measure of the ingress and egress durations. The ingress durations obtained from the three {\it RXTE}--ASM profiles, 
three {\it Swift}--BAT profiles and the {\it MAXI}--GSC profiles are 0.08, 0.1, 0.14, 0.04, 0.05, 0.04, 0.07 of the orbital period while the egress 
durations are 0.04, 0.04, 0.04, 0.03, 0.04, 0.04, 0.06 respectively. There is better overall agreement between the eclipse egress durations compared 
to the ingress durations, the latter seems to be larger in the {\it RXTE}--ASM and {\it MAXI}--GSC data. This is in agreement with additional absorption 
before the eclipse due to the presence of an accretion wake \citep{coley2015}, effect of which is more prominent in the ASM energy band. However, 
for the main purpose of this paper, we require the mid-eclipse time and duration of the total eclipse. By using the double ramp function given in 
Equation ~\ref{ramp}, these two quantities are measured independent of the ingress and egress durations and their variation or energy dependence. \\
We have also used {\it RXTE}--PCA observations (ObsId 30094), from April 1999 to August 1999, covering 
eclipse ingress and egress for 120 days or 11 orbital cycles. The light-curve of {\it RXTE}--PCA observations is folded with the orbital period estimated for the first 
segment of {\it RXTE}--ASM light curve. 
\par
Figure \ref{asm} shows the orbital intensity profiles near the eclipses created from 3 light-curve segments of {\it RXTE}--ASM, 3 light-curve segments 
of {\it Swift}--BAT and the {\it MAXI}--GSC light-curve, fitted with a two ramp function. 
The blue vertical line in each panel denotes 
the measured mid-eclipse phase, red line denotes the mid-eclipse phase expected from the orbital period decay extrapolated from \cite{rubin1996} and the green line 
is the expected mid-eclipse phase from the orbital period decay extrapolated from \cite{falanga2015}. 
All the mid-eclipse times extrapolated from \cite{rubin1996} and all but one from \cite{falanga2015} appear before the measured mid-eclipse phase for all the orbital 
intensity profiles, implying that the orbital period decay is lower than the previously estimated values. 
\par
The mid-eclipse times used for the analysis, both the earlier reported values as well as the new mid-eclipse times, are given in Table 1. The previously 
quoted uncertainities on the mid-eclipse times from {\it Uhuru}, {\it Copernicus} and {\it EXOSAT} were re-estimated by \cite{rubin1996} and were calculated disparately 
from that of the mid-eclipse times from {\it Granat}/Watch and {\it BATSE} observations given in Table 2 in \cite{rubin1996}. The different error bars quoted previously 
are only the statistical errors. However, eclipse measurements in different energy bands with different instruments are also likely to have some systematic 
differences. To have a more realistic estimates of error bars and only with the aim of measuring a long term period derivative, we have 
re-estimated the errors on the mid-eclipse times, which are now likely to include the systematic differences between the different instruments. We have 
divided the mid-eclipse times into two segments, from MJD: 41452 - 49150 and MJD: 49151 - 56468. A linear fit is done separately on these two segments 
and the average standard deviation of the data-points from the linear fit is taken as the errors on the mid-eclipse times for these two segments 
(given in Table 1). 

\subsection{Systematic errors associated with the mid-eclipse time measurement with {\it EXOSAT}--GSPC and {\it RXTE}--PCA}
As seen in top panel of Figure \ref{gspc_pca}, the orbital intensity profile of 4U 1700--37 constructed out of long term light curves from the X-ray all sky 
monitors like {\it Swift}--BAT show a smooth profile with a sharp eclipse due to averaging effects of observations made over many orbits. However at short 
timescales, the light curves show strong variations, including flares and dips, characteristic of many HMXBs. Among the light curves of 4U 1700--37, 
those with single observation covering eclipse like {\it EXOSAT}--GSPC and {\it RXTE}--PCA show strong flares and dips (bottom panel of Figure \ref{gspc_pca}), 
which introduces significant uncertainities in the eclipse ingress and egress profiles and hence, in the mid-eclipse time determination. The mid-eclipse time 
estimated from the {\it EXOSAT} observation in 1985 was quoted a statistical error of 0.003 days \citep{haberl1989}. We have extracted the 
{\it EXOSAT}--GSPC light curve in the 8--14 keV band and constructed the orbital intensity profile by folding it with the orbital period mentioned 
in \cite{haberl1989}. As seen in the right panel of Figure \ref{gspc_pca}, the determination of the exact point of ingress and egress of the eclipse is 
complicated by the presence of flares and/or dips around the eclipse ingress and egress. To further emphasize on the contribution of flares in 
uncertainities on the mid-eclipse times in case of pointed observations, we have overlaid the orbital intensity profile from the {\it RXTE}--PCA 
observations (ObsId:30094) in the same plot. From the right panel of Figure \ref{gspc_pca},  we infer that the statistical error of 0.003 days or 260 secs 
quoted in \cite{haberl1989} is an underestimate of the actual error on the mid-eclipse time, as there is a larger systematic error owing to flares/dips. 
Determination of eclipse duration is also complicated by the presence of same flares and/or dips. In fact the eclipse durations determined from single 
observations with {\it Uhuru} \citep{jones1973}, {\it Copernicus} \citep{branduardi1978} and {\it EXOSAT} \citep{haberl1989} are considerably larger with 
large error bars than those estimated from long term light curves from the all sky monitors.
\par
We have re-calculated the errors on the {\it EXOSAT} mid-eclipse time as mentioned in Section 2.1. It appears that the faster rate of orbital period decay 
in the earlier estimates \citep{rubin1996,falanga2015} were result of small statistical error considered for the {\it EXOSAT} mid-eclipse data. 

\subsection{Orbital evolution of 4U 1700--37}

The mid-eclipse times given in Table 1 along with their errors are fitted to a quadratic function
\begin{equation}
\label{fit}
 T_{N} = T_{0} + PN +\frac{1}{2} P{\dot{P}}N^{2}
\end{equation}
where $T_N$ is the mid-eclipse time of the {\it N}$^{th}$ orbital cycle. {\it P} is the orbital period in days and ${\dot{P}}$ is the orbital period 
derivative, both at time $\mathrm{T_{0}}$. 
\par
The definition of $\mathrm{T_{0}}$ and orbit number are same as mentioned in \cite{rubin1996}.
The best fit to the mid-eclipse times for a constant orbital period gives $\mathrm{T_{0}} = 49149.412 \pm 0.006$ MJD,  orbital period 
{\it P} = 3.411660 $\pm$ 0.000004 days with a $\chi^{2} = 28.3$ for 22 degrees of freedom. The mid-eclipse times fitted with a quadratic function as in Equation \ref{fit}
gives the values of orbital period decay ${\dot{P}}/P = -(4.7 \pm 1.9) \times 10^{-7}$ yr$^{-1}$ with a $\chi^{2} = 23.8$ for 21 degrees of freedom. 
The orbital ephemeris 
is also tabulated in Table 2. For the orbital ephemeris given in \cite{falanga2015}, the value of $\chi^{2} = 67.15$ for 24 degrees of freedom. 
The orbital period decay rate of 4U 1700--37 determined here is smaller compared to the earlier estimates \citep{haberl1989,rubin1996,falanga2015}. 
A plot of the delay in mid-eclipse times with respect to a constant orbital period as a function of the number of orbital cycles, along with the best 
fit quadratic function is given in Figure \ref{delay1}. The quadratic function reported in \cite{rubin1996} and \cite{falanga2015} are overlaid on the 
same plot. 

\begin{figure*}
\centering
\includegraphics[scale=0.25]{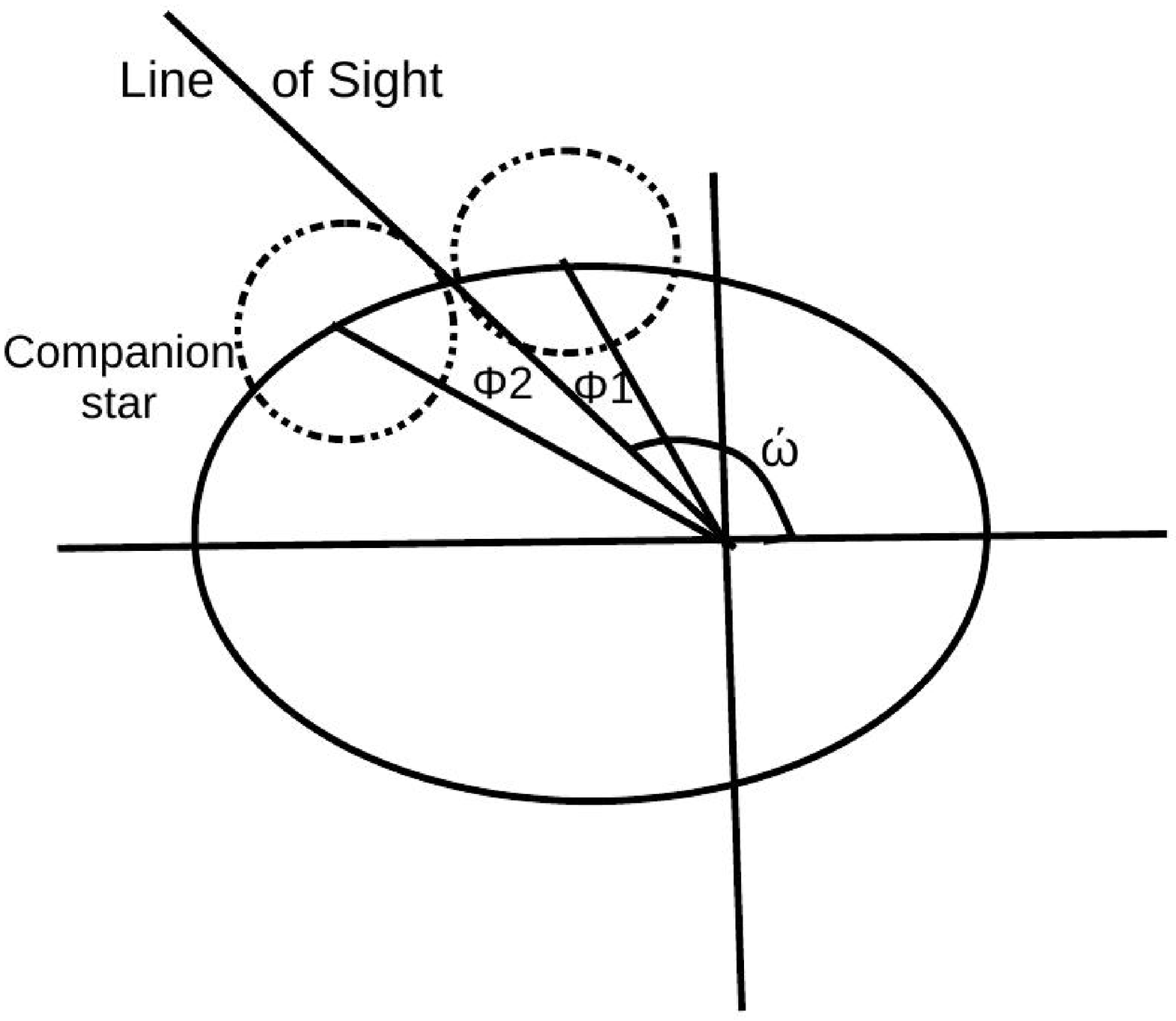} 
\caption{The ellipse represents the motion of the companion star with respect to the compact star at the focus. See text for description of the various angles.}
\label{ellipse}
\end{figure*}

\begin{figure*}
\centering
\includegraphics[angle=-90,scale=0.45]{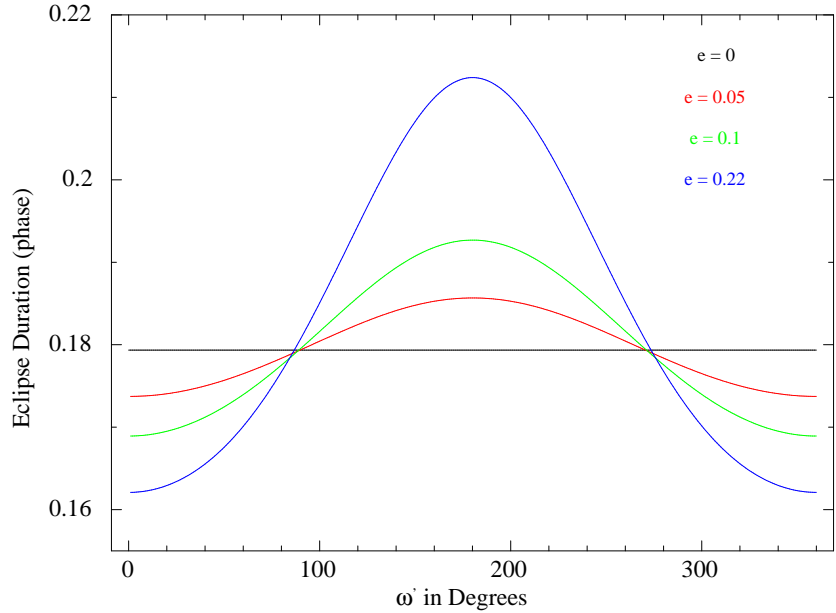} 
\caption{Plot of variation of eclipse duration as a function of $\omega^{'}$ for different value of eccentricity, for an orbital inclination of 
66$^{\circ}$. Increasing values of eccentricity increases the ratio of maximum to minimum eclipse duration.}
\label{eccentricity}
\end{figure*}

\section{Eccentricity and Apsidal Motion of 4U 1700--37} 

In a close binary stellar system, the rate of apsidal motion due to tidal forces is given by \citep{claret1993}

\begin{equation}
\label{stellar_constant}
 \frac{\dot \omega}{360} = k (\frac{R_{\star}}{a})^{5} (15qg(e) + \Omega^{2}(1+q)f(e)) \quad \mathrm{deg/cycle} \\
\end{equation}
where {\it e} is the eccentricity, R$_{\star}$ is the companion star radius, {\it a} is the binary separation, {\it q} is the mass ratio of the compact 
object to the companion star and $\Omega$ is ratio of the rotational velocity of the companion star to its orbital angular velocity
$$g(e) = (1 + \frac{3}{2} e^{2} + \frac{1}{8} e^{4}) (1-e^{2})^{-5}$$ 
$$f(e) = (1 - e^{2})^{-2} $$
Using the binary parameters of 4U 1700--37 along with its uncertainities (R$_{\star}$, {\it a}, {\it q}, P$_{orb}$ from Table 3), a reasonable value of 
stellar constant {\it log k} of $-2.2$ corresponding to the companion star HD 153919 type \citep{claret2004} and {\it e} in the range of 0.01--0.22, 
we get an apsidal motion rate of $10 \pm 3$ degrees/year. For Vela X--1 and 4U 1538--522, using the binary parameters from Table 7 in \cite{falanga2015}, 
an estimation of the rate of apsidal motion is $\sim$ 1 degree/yr and 5 degree/yr respectively, similar to that measured in these sources \citep{deeter1987,falanga2015}. 
The major source of uncertainity in estimating the rate of apsidal motion arises from the value of stellar constant {\it k}, which for some HMXBs are constrained by 
observations of apsidal motion in X-ray binaries \citep{raichur2010}.
\par
The two interesting consequences of a large apsidal advance rate in an eclipsing X-ray binary are that the delay in the mid-eclipse times and the value of 
eclipse duration both would vary with a period of 360$^{\circ}/\dot\omega$. The delay in mid-eclipse times due to apsidal motion is seen in close 
eclipsing optical binary stars \citep{gimenez1983,wolf2004,zasche2014}. The mid-eclipse times (Table 1) and the corresponding 
eclipse duration measurements of 4U 1700--37 have been carried out with a large number of instruments of different sensitivities and in different energy 
bands. We have mentioned before that this causes significant systematic differences and therefore are not ideal to investigate the effects of 
apsidal motion. Eclipse duration is also found to be dependent on the energies of the X-rays with some eclipses lasting longer at lower energies than 
higher energies \citep{meer2007}. So we have used hard X-ray light curve from {\it Swift}-BAT which has the highest statistical quality (second panel in Figure 1) 
and have divided it into 10 segments of 1 year data and searched for the signatures of an apsidal motion for an eccentric orbit. The yearly measurements of the eclipse data 
(mid-eclipse time and eclipse duration) are useful to probe an $\dot\omega$ in the range of 5-200 degrees per year. Since these 10 measurements are from the 
data obtained with the same instrument, it is unlikely to have much systematic difference between different data points.

\subsection{Mid-eclipse time variation due to apsidal motion}

For a system having a small eccentricity and undergoing an apsidal motion, the mid-eclipse times show a sinusoidal 
pattern given by the following Equation \citep{gimenez1983}

\begin{equation}
 \label{apsidal_truncate_eqn}
 T_{N} - (T_{0} + PN) = \frac{eP_{a}}{\pi} \mathrm{cos}(\omega_{0} + \Delta \omega N)
\end{equation}
where $\Delta \omega$ is the change in $\omega$ in one orbital cycle. P$_{a}$ is the anomalistic orbital period defined by the interval of time between two consecutive periastron 
passages and given by :
\begin{equation}
  P_{a} = \frac{P}{(1- \frac{\Delta \omega}{2\pi})} \nonumber
\end{equation}
For moderate values of $\Delta\omega$, P$_{a} \sim$ P (orbital period).
\par
The orbital inclination of the binary system 4U 1700--37 is of the order of 60$^{\circ}$--70$^{\circ}$ \citep{rubin1996,falanga2015} and will lead to a 
change in the estimation of $\omega_{0}$ of the system. However, the estimation of $\Delta \omega$ and eccentricity {\it e} is independent of the 
orbital inclination of the system.

\subsection{Variation in eclipse duration due to apsidal motion}

By considering the motion of the companion star around the compact object, we initially calculated the eclipse durations as a function of the angle of periastron for 
different values of {\it e} for an orbital inclination of 90$^{\circ}$. In Figure \ref{ellipse}, the ellipse represents the motion of the companion star 
with respect to the compact star at the centre. $\phi_1$ and $\phi_2$ represent the eclipse ingress and egress, $R_{\star}$ is the companion star radius 
and $\omega^{'}$ is the angle between the periastron and the line of sight.
\begin{equation}
\label{omega}
 \omega^{'} = \omega - \frac{\pi}{2}
\end{equation}

If $\theta$ is the position angle of the companion star, the eclipse ingress and
egress ($\phi_{1,2}$) can be determined from
\begin{equation}
 \mathrm{sin} \phi_{1,2} = \beta (1 + e \mathrm{cos}\theta_{1,2})
\end{equation}
where $\beta = \frac{R_{\star}}{a(1-e^{2})}$.

The eclipse duration can be estimated as:

\begin{equation}
\label{duration}
\Delta T(\omega^{'}) = \frac{a^{2}(1-e^{2})^{2}}{L} \int_{\omega^{'} - \phi_1}^{\omega^{'} + \phi_2} \frac{d\theta}{(1 + e \mathrm{cos} \theta)^{2}}
\end{equation}

where $L = (G(M_{\star}+M_{C})a(1-e^2))^{\frac{1}{2}}$.

\par
Inclusion of the orbital inclination of the system would lead to a change in the projected 
semi-major axis $a_{x}${\it sin i} instead of {\it a} in Equation \ref{duration}. 
\begin{equation}
\Delta T(\omega^{'}) = \frac{(a_{x}sini)^{2}(1-e^{2})^{2}}{L} \int_{\omega^{'} - \phi_1}^{\omega^{'} + \phi_2} \frac{d\theta}{(1 + e \mathrm{cos} \theta)^{2}}
\end{equation}

However, the ratio of maximum value of the eclipse duration to its minimum value will 
remain the same, which is used to estimate the eccentricity of the system. Using the values of the binary parameters ($M_{\star}, M_{C}, i, \mathrm{and} ~a$ from Table 3), 
we have calculated the eclipse duration as a function of $\omega^{'}$ for different values of eccentricity, shown in Figure \ref{eccentricity}. 
The maximum value of eclipse duration to its minimum value is approximately equal to $\frac{1 + e}{1 - e}$. 

\begin{figure*}
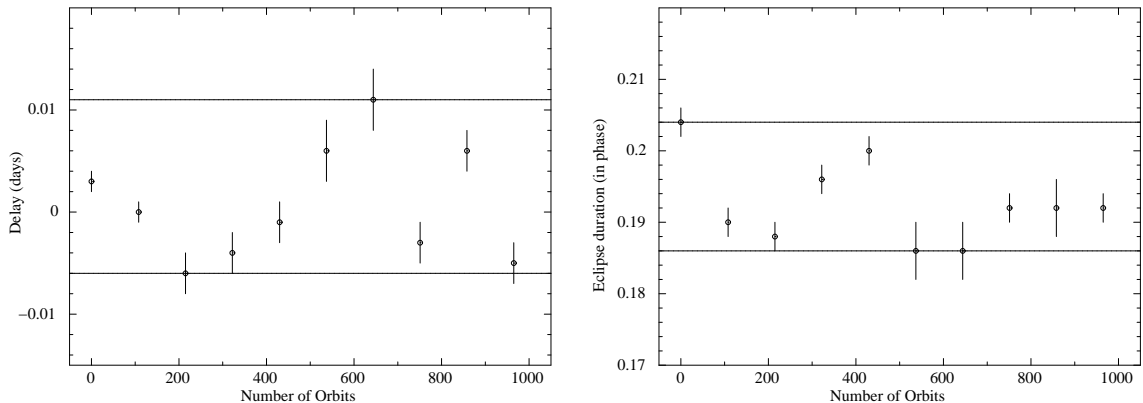

\centering
\includegraphics[angle=-90,scale=0.3]{swiftbat_delay.ps} 
\includegraphics[angle=-90,scale=0.3]{swiftbat_duration.ps} 
\caption{{\it Left panel}: Delay in mid-eclipse times with respect to a constant orbital period for 10 segments of 1 year {\it Swift}--BAT light curves 
plotted as function of number of orbits, along with solid lines showing the maximum and minimum value of the delay. This limit is compared to the amplitude 
($\frac{eP}{\pi}$) in Equation \ref{apsidal_truncate_eqn} to give an eccentricity value of 0.008. {\it Right panel}: Plot of eclipse duration as function 
of orbit number calculated from 10 segments of 1 year {\it Swift}-BAT light curves, along with solid lines showing the maximum and minimum value of the 
eclipse duration. This limit is compared to $\frac{1 + e}{1 - e}$ to give an eccentricity value of 0.05.}
\label{swift}
\end{figure*}

\subsection{The {\it Swift}--BAT eclipse data}

The mid-eclipse time measurements from the 10 segments of 1 year {\it Swift}--BAT light curve are shown in the left panel of Figure \ref{swift} after 
subtracting the linear component. It shows a maximum variation of about $\pm$ 0.01 days. The eclipse durations measured from the same time segments 
obtained by fitting the two ramp function as described in Section 2.1 are shown in right panel of Figure \ref{swift}. The maximum variation in eclipse 
duration is less than $\pm$ 5$\%$. These two data sets, of mid-eclipse time variation and variation of eclipse duration also given in Table 4, do not show 
any periodic variation with a same period. It is therefore not possible to determine any apsidal motion rate from these data. However, the maximum variation 
in both the plots can be used to put upper limits on the eccentricity by comparing them with the amplitude term ($\frac{eP}{\pi}$) in 
Equation \ref{apsidal_truncate_eqn} and to $\frac{1 + e}{1 - e}$ respectively.
\par
The upper limit on eccentricity of 4U 1700--37, obtained from the two methods are 0.008 and 0.05 respectively. We note here that these limits are 
applicable for apsidal motion rate greater than about 5 degrees per year. The limits are much smaller than the eccentricity reported from Doppler 
velocity measurements of the companion star by \cite{hammerschlag2003}.

\section{Discussions}

\subsection{Possible Causes of Orbital Period Decay}
Orbital period changes are found to occur in High Mass X-ray binaries like Cen X--3, SMC X--1 \citep{raichur2010}, LMC X--4 \citep{naik2004}, 
OAO 1657--415 \citep{jenke2012} and in Low Mass X-ray binaries like Her X--1 (Staubert et al. 2009), EXO 0748--676 \citep{wolff2009}, 4U 1822--37 
\citep{jain2010}, and SAX J1808.4--3658 \citep{jain2008}. In case of Low Mass X-ray binaries, the orbital evolution is assumed to occur mainly due to 
conservative mass transfer from the companion star to the neutron star or due to mass loss from disk winds. 
\par
In case of High Mass X-ray binaries, orbital period decay occurs due to stellar wind driven angular momentum loss and/or strong tidal interactions 
between the binary components. Tidal interactions between binary components of Cen X--3, LMC X--4 and SMC X--1 are the primary cause of orbital period 
evolution because these are the short orbital period HMXBs with strong tidal effects and the mass loss rate of $10^{-7}-10^{-6}$ M$_{\odot}$ yr$^{-1}$ in 
these sources is not sufficient to account for the orbital period decay due to wind driven angular momentum loss \citep{kelley1983,levine1993,levine2000}. 
The orbital period decay estimated in these systems are of the order (0.9 -- 3.4) $\times 10^{-6}$ yr$^{-1}$ \citep{falanga2015}. The orbital period decay 
seen in 4U 1700--37 is smaller than that seen in these systems, inspite of it having the largest (R$_{\star}$/a) ratio compared to the other binaries. On the other hand, 
the orbital period decay seen in OAO 1657--415, which has a larger orbital period of 10.44 days \citep{chakrabarty1993}, can be explained with wind 
driven angular momentum loss \citep{jenke2012}. 
In case of 4U 1700--37, the earlier estimate of ${\dot{P}}/P = -3 \times 10^{-6}$ yr$^{-1}$ by \cite{rubin1996} was accounted by wind driven angular 
momentum loss. As mentioned in \cite{rubin1996}, by taking into account the uncertainities in various factors contributing to orbital period decay due to 
stellar wind driven angular momentum loss, the mass loss rate can be as less as 10\% of the total and the present estimated orbital period decay could 
be solely driven by it. It would be interesting to investigate the models evaluating the contribution of stellar wind driven angular momentum loss and 
tidal interactions in the orbital decay rate seen for this binary system \citep{lecar1976,hut1981,klis1984,brookshaw1993,heuvel1994}.

\subsection{Eccentricity of the binary orbit}

The upper limit on eccentricity of the orbit of 4U 1700--37 put from the limits of residuals in the mid-eclipse times and limits on variation in the 
eclipse duration is quite low; {\it e} $\sim$ 0.008 and 0.05 respectively. This is in contrast with $e \sim 0.22$ from the radial velocity 
measurements with {\it IUE} data \citep{hammerschlag2003}. In the presence of a significant apsidal motion, the radial velocity measurements at different 
orbital phases (with respect to the mid-eclipse times) in data spread over several years can not be put together in a simple way. The {\it IUE} data from 
which an eccentricity of 0.22 was reported are not sampled densely enough for a joint fit to measure eccentricity. The present work with {\it Swift}-BAT 
(Section 3.3) indicates a nearly circular orbit for this system if the apsidal motion rate is in the range of 5-200 degrees per year. An even higher rate 
of apsidal motion along with a significant eccentricity can be ruled out from the fact that edges of the eclipse ingress and egress are quite sharp. 
In the presence of a large apsidal motion rate and eccentricity, the edges of the eclipse profiles with {\it Swift}--BAT shown in the second panel of 
Figure \ref{asm} would be smoothed out. 

Measurement of correct orbital parameters of the system would require new radial velocity measurements with good orbital 
coverage in a single epoch. In addition, the LAXPC instrument of recently launched mission {\it ASTROSAT} \citep{paul2013} will either 
detect or lower the upper limit of pulse fraction of 4U 1700--37 in a wide X-ray energy band of 3-80 keV. We note here that the accurate determination of the 
orbital parameters of this source and hence the mass of the compact object, is of high interest as it is either a very high mass neutron star or a very 
low mass black hole \citep{clark2002}.

\vspace{15mm}
\textbf{ACKNOWLEDGEMENT}\\
We are very thankful to the referee for careful reading of the manuscipt and for making suggestions which have improved the paper.
The data used for this work has been obtained through the High Energy Astrophysics Science Archive (HEASARC) 
Online Service provided by NASA/GSFC. This research has made use of public light-curves from {\it Swift}-BAT site and {\it MAXI} data provided by RIKEN, 
JAXA and the MAXI team. The authors thank S. Sridhar and A. Gopakumar for useful discussions.

 \begin{sidewaystable*}
\footnotesize
\centering
\label{table1}
\caption{Mid-eclipse time measurements along with errors calculated as mentioned in Section 2.1}
\begin{tabular}{| c c c c c c |}
\hline
Satellite          & Energy Range & Mid-eclipse Time (MJD) & Mid-eclipse Time (MJD)       & Mid-eclipse Time (MJD)        & Reference             \\
                   &              & with reported errors   & with re-calculated errors    & with statistical errors       &                       \\
\hline
\vspace{3mm}
{\it Uhuru}$^{\star}$        & 2-6 keV      & 41452.64(1)            & 41452.640(34)                & -                             & \cite{rubin1996}    \\
\vspace{3mm}
{\it Copernicus}   & 2.8-8.7 keV  & -                      & 42230.625(34)                & -                             & \cite{mason1976}     \\
\vspace{3mm}
{\it Copernicus}$^{\star}$   & 2.8-8.7 keV  & 42609.25(1)            & 42609.250(34)                & -                             & \cite{branduardi1978} \\
\vspace{3mm}
{\it Copernicus}$^{\star}$   & 2.8-8.7 keV  & 42612.646(10)          & 42612.646(34)                & -                             & \cite{branduardi1978}  \\
\vspace{3mm}
{\it Copernicus}$^{\star}$   & 2.8-8.7 keV  & 43001.604(10)          & 43001.604(34)                & -                             & \cite{branduardi1978}  \\
\vspace{3mm}
{\it Copernicus}$^{\star}$   & 2.8-8.7 keV  & 43005.000(10)          & 43005.000(34)                & -                             & \cite{branduardi1978}   \\
\vspace{3mm}
{\it EXOSAT}       & 2-10 keV     & 46160.840(3)           & 46160.840(34)                & -                             & \cite{haberl1989}      \\
\vspace{3mm}
{\it Granat}/WATCH$^{\star}$ & 8-20 keV     & 48722.940(31)          & 48722.940(34)                & -                             & \cite{sazonov1993}     \\
\vspace{3mm}
{\it BATSE}        & 20-120 keV   & 48651.365(31)          & 48651.365(34)                & -                             & \cite{rubin1996}       \\
\vspace{3mm} 
{\it BATSE}        & 20-120 keV   & 49149.425(27)          & 49149.425(34)                & -                             & \cite{rubin1996}       \\
\vspace{3mm}
{\it INTEGRAL}     & 17-40 keV    & 52861.29(2)            & 52861.290(17)                & -                             & \cite{falanga2015}      \\
\vspace{3mm}
{\it INTEGRAL}     & 17-40 keV    & 53270.69(2)            & 53270.690(17)                & -                             & \cite{falanga2015}   \\
\vspace{3mm}
{\it INTEGRAL}     & 17-40 keV    & 53472.00(2)            & 53472.000(17)                & -                             & \cite{falanga2015}   \\
\vspace{3mm}
{\it INTEGRAL}     & 17-40 keV    & 53785.82(3)            & 53785.820(17)                & -                             & \cite{falanga2015}   \\
\vspace{3mm}
{\it INTEGRAL}     & 17-40 keV    & 54164.55(2)            & 54164.550(17)                & -                             & \cite{falanga2015}   \\
\vspace{3mm}
{\it INTEGRAL}     & 17-40 keV    & 54341.97(3)            & 54341.970(17)                & -                             & \cite{falanga2015}   \\
\vspace{3mm}
{\it RXTE}--ASM    & 5-12 keV     &   -                    & 50865.485(17)                & 50865.485(8)                  & Present Work         \\
\vspace{3mm}
{\it RXTE}--ASM    & 5-12 keV     &   -                    & 52445.102(17)                & 52445.102(11)                 & Present Work        \\
\vspace{3mm}
{\it RXTE}--ASM    & 5-12 keV     &   -                    & 54533.033(17)                & 54533.033(13)                 & Present Work           \\
\vspace{3mm}
{\it Swift}--BAT   & 15-50 keV    &   -                    & 54028.076(17)                & 54028.076(1)                  & Present Work     \\
\vspace{3mm}
{\it Swift}--BAT   & 15-50 keV    &   -                    & 55246.031(17)                & 55246.031(2)                  & Present Work          \\
\vspace{3mm}
{\it Swift}--BAT   & 15-50 keV    &   -                    & 56467.395(17)                & 56467.395(1)                  & Present Work         \\
\vspace{3mm}
{\it MAXI}--GSC    & 5-20 keV     &   -                    & 55935.181(17)                & 55935.181(9)                  & Present Work           \\
\vspace{3mm}
{\it RXTE}--PCA    & 2-60 keV     &   -                    & 51295.331(17)                & -                             & Present Work          \\

\hline
\end{tabular}
\footnotetext{$\star$ denotes the mid-eclipse times whose errors were re-analysed by \cite{rubin1996}.}
\end{sidewaystable*}

\begin{table*}
\centering
\label{par}
\caption{Orbital ephemeris of 4U 1700--37 estimated in Section 2.3}
\begin{tabular}{| c c c |}
\hline
\hline
Epoch                          &  T$_{0}$	                       & 49149.412 $\pm$ 0.006 MJD \\
Orbital period                 &  P$_{orb}$                            & 3.411660 $\pm$ 0.000004 days \\
Orbital period decay           &  $\dot{P}/P$                          & -$(4.7 \pm 1.9) \times 10^{-7}$ yr$^{-1}$\\
\hline
\end{tabular}
\end{table*}

\begin{table*}
\centering
\label{table2}
\caption{ Stellar parameters for HD 153919 and orbital parameters of 4U 1700--37}
\begin{tabular}{| c c c |}
\hline
Parameter                      &  Value                                         & Reference \\
\hline
R$_{\star}$	               & 22$\pm 2$ R$_{\odot}$                 & \cite{falanga2015} \\
M$_{\star}$                    & 46 $\pm$ 5 M$_{\odot}$                & \cite{falanga2015} \\
M$_{C}$                        & 1.96 $\pm$ 0.19 M$_{\odot}$           & \cite{falanga2015} \\
$\Omega$                       & 0.47                                  & \cite{falanga2015} \\
{\it log k}                    & -2.2                                  & \cite{claret2004} \\
{\it i}                        & 66$^{\circ}$                          & \cite{rubin1996} \\
{\it a}                        & 35 R$_{\odot}$                        & \cite{falanga2015} \\
P$_{orb}$                      & 3.4117 days                           & \\
\hline
\end{tabular}
 \end{table*}

\begin{table*}
\centering
\label{table3}
\caption{Mid-eclipse time and eclipse duration measurements along with errors (1 $\sigma$ confidence limit) for 10 segments of 1 year each {\it Swift}--BAT data}
\begin{tabular}{| c c c |}
\hline
Experiment & Mid-eclipse Time (MJD) & Eclipse Duration (in Phase) \\
\hline
\vspace{3mm}
Segment 1  & 53601.624$\pm$0.001    & 0.204$\pm$0.002    \\
\vspace{3mm}
Segment 2  & 53970.078$\pm$0.001    & 0.190$\pm$0.002    \\
\vspace{3mm}
Segment 3  & 54335.115$\pm$0.002    & 0.188$\pm$0.002     \\
\vspace{3mm}
Segment 4  & 54700.164$\pm$0.002    & 0.196$\pm$0.002     \\
\vspace{3mm}
Segment 5  & 55068.624$\pm$0.002    & 0.200$\pm$0.002      \\
\vspace{3mm}
Segment 6  & 55433.678$\pm$0.003    & 0.186$\pm$0.004      \\
\vspace{3mm}
Segment 7  & 55798.726$\pm$0.003    & 0.186$\pm$0.004       \\
\vspace{3mm}
Segment 8  & 56163.759$\pm$0.002    & 0.192$\pm$0.002        \\
\vspace{3mm}
Segment 9  & 56528.811$\pm$0.002    & 0.192$\pm$0.004        \\
\vspace{3mm}
Segment 10 & 56893.847$\pm$0.002    & 0.192$\pm$0.002         \\
\hline
\end{tabular}
\end{table*}

\end{document}